\documentclass[aps,amsmath,amssymb, prl,twocolumn,superscriptaddress,showpacs]{revtex4-1}

\usepackage{graphicx}
\usepackage{dcolumn}
\usepackage{bm}
\usepackage[utf8]{inputenc}
\usepackage[ngerman,english]{babel}
\usepackage{multirow}

\begin{document}


\title{Origin of Low-Lying Enhanced $E1$ Strength in Rare-Earth Nuclei}


\author{M. Spieker}
\email[]{spieker@ikp.uni-koeln.de}
\affiliation{Institut f\"{u}r Kernphysik, Universit\"{a}t zu K\"{o}ln, Z\"{u}lpicher Straße 77, D-50937 K\"{o}ln, Germany}

\author{S. Pascu}
\affiliation{Institut f\"{u}r Kernphysik, Universit\"{a}t zu K\"{o}ln, Z\"{u}lpicher Straße 77, D-50937 K\"{o}ln, Germany}
\affiliation{National Institute for Physics and Nuclear Engineering, R-77125, Bucharest-Magurele, Romania}

\author{A. Zilges}
\affiliation{Institut f\"{u}r Kernphysik, Universit\"{a}t zu K\"{o}ln, Z\"{u}lpicher Straße 77, D-50937 K\"{o}ln, Germany}

\author{F. Iachello}
\affiliation{Center for Theoretical Physics, Sloane Physics Laboratory, Yale University, New Haven, Connecticut 06520-8120, USA}


\date{\today}

\begin{abstract}

The experimental {\it E1} strength distribution below 4 MeV in rare-earth nuclei suggests a local breaking of isospin symmetry. In addition to the octupole states, additional $J^{\pi}$=~1$^{-}$ states with enhanced $E1$ strength have been observed in rare-earth nuclei by means of ($\gamma$,$\gamma$') experiments. By reproducing the experimental results, the {\it spdf} interacting boson model calculations provide further evidence for the formation of an $\alpha$ cluster in medium-mass nuclei and might provide a new understanding of the origin of low-lying $E1$ strength.

\end{abstract}

\pacs{21.10.Re, 21.60.Ev, 21.60.Gx, 25.20.Dc}

\maketitle

\begin{figure}[b]
\centering
\includegraphics[width=.8\linewidth]{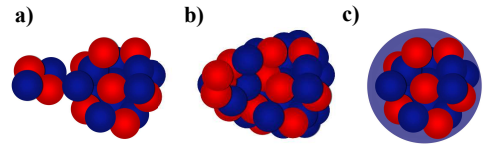}
\caption{\label{fig:e1shapes} (Color online) Macroscopic interpretations of the different low-lying dipole modes. {\bf (a)} $\alpha$-clustering mode, {\bf (b)} octupole mode, and {\bf (c)} neutron-skin oscillation (pygmy dipole resonance (PDR)).}
\end{figure}

The atomic nucleus is a unique laboratory which allows us to study how matter is built from the smallest scales ($10^{-15}\,\mathrm{m}$) to stellar scales. In this mesoscopic system the symmetry between protons and neutrons, i.e. global isospin symmetry~\cite{Warn06a}, is a fundamental assumption. However, already in 1985 one of us (F. Iachello) proposed that even at lower excitation energies local rather than global isospin symmetry is realized and that enhanced ${E1}$ transitions could possibly further test the mechanisms by which this local symmetry is broken in atomic nuclei~\cite{Iach85a}. Below the particle-emission threshold, two kinds of {\it E1} excitations have been intensively studied during the last decade, the octupole modes~\cite{Butl96a, Knei06a} and the pygmy dipole resonance (PDR)~\cite{Sav13a}, see Figs.~\ref{fig:e1shapes}\,{\bf b)} and \ref{fig:e1shapes}\,{\bf c)}. For both modes, it has been pointed out that they could have an important impact on fundamental physical properties, {\it e.g.}, on enhancing the sensitivity for permanent electric dipole moment measurements in the case of the octupole mode~\cite{Gaff13a} and on the equation of state in the case of the PDR~\cite{Pieka12a}. 

In both cases, the nonuniform distribution of protons and neutrons in simplified macroscopic models gives rise to enhanced {\it E1} transitions at energies below the neutron separation energy. A clustering mode has also been proposed~by one of us~\cite{Iach85a} where the $\alpha$ cluster might be considered as the simplest and energetically most favored realization, see Fig.~\ref{fig:e1shapes}\,{\bf a)}. This mode would give rise to enhanced $E1$ transitions due to the oscillation of the $\alpha$ cluster relative to the remaining bulk~\cite{Iach82a, Iach85a}. 

In physics, clustering phenomena are of interest in many fields. For example on the microscopic scale ultracold gases in traps~\cite{Wen09a}, the electron-hole-exciton system in excited semiconductors~\cite{mosk00a, Alm14a} or cluster systems for quantum computation~\cite{Brie01a, Walt05a} are studied and on the macroscopic scale, {\it e.g.}, clusters of stars~\cite{Li14a} and galaxies~\cite{Hand12a, Stani09a} are of recent interest. In nuclear physics $\alpha$ clustering is a well-established feature of lighter nuclei~\cite{vOertz06a, sotancp3}, {\it e.g.}, $^{12}$C~\cite{Bij00a, Mar14a, Toh01a} and $^{16}$O~\citep{Bij14a, Ito14a, Toh01a}, and its implications for the $E1$ strength have been discussed~\cite{Alh82a, Gai83a, He14a}. First strong indications of a $^{208}$Pb+$\alpha$ system have been observed by means of enhanced $E1$ transitions between excited states of $^{212}$Po~\cite{Ast10a} and also the possibility of $\alpha$-cluster states in the actinides was discussed long ago, {\it e.g.}, Refs.~\cite{Iach82a, Dal83a, Gai83b, Gai88a}. Recently, an exploratory calculation for $^{212}$Po was presented~\cite{Roe14a}, indicating the existence of $^{208}$Pb+$\alpha$ configurations when four-particle correlations are added to the shell-model calculations. This calculation provided a first hint at how to extend the well-established Tohsaki-Horiuchi-Schuck-R{\"o}pke wave function concept used for $\alpha$-like condensates in light nuclei~\cite{Roe98a, Toh01a, Sog09a, Zho13a}  to heavier nuclei. However, the general existence of $\alpha$ clustering in heavier nuclei remains an open question. Since $\alpha$ clustering in nuclei provides interesting insights about the formation of bosonic clusters in strongly coupled fermionic systems, indentifying new signatures of $\alpha$ clustering in heavier nuclei is of general scientific interest.

In ${}^{150}$Nd strong reduced $\alpha$ widths were observed for a group of states with $J^{\pi}$=~0$^+$, 2$^+$, and 4$^+$ between 2 and 2.5~MeV in the $(d,{}^{6}Li)$ reaction, which was interpreted as new evidence for a rotational band built upon an $\alpha$-cluster state~\cite{Jaen82a}. These results triggered systematic ($\gamma$,$\gamma$') studies to find candidates for the expected $J^{\pi}$=~1$^-$ states of the $\alpha$-clustering mode~\cite{Iach85a}. Indeed, several enhanced {\it E1} transitions were observed in ${}^{142-150}$Nd~\cite{Pitz90a, Fried92a, Eck97a}, ${}^{148-154}$Sm~\cite{Zieg93a}, ${}^{156-160}$Gd~\cite{Pitz89a, Fried94a}, ${}^{162}$Dy~\cite{Fried92a}, and other rare-earth nuclei~\cite{Zil91a, Frans98a} below 4\,MeV. First attempts to qualitatively describe the experimentally observed peculiar increase of the $E1$ strength connected with the proposed quadrupole-octupole coupled (QOC) $1^-_1$ state~\cite{Knei06a} towards shell closures~\cite{Frans98a} in a microscopic manner were presented in Refs.~\cite{Zamf90a, Zil92a, Cott94a, Heyd97a, Andrej01a, Grin94a, Ponom98a}. There, it was proposed that one-particle-one-hole ($1p-1h$) admixtures, originating from the tail of the electric giant dipole resonance (GDR), into the QOC state could explain the enhanced $E1$ strength and that destructive interference of both components causes the decrease when moving away from the shell closure. Modern energy-density functional (EDF) calculations of $E1$ strength in transitional rare-earth nuclei were performed in Ref.~\cite{Yosh11a}. However, the authors pointed out that the study of multiphonon states, e.g., $2^+ \otimes 3^-$, with the EDF method is challenging and these have consequently only been studied in the spherical semimagic $N=82$ isotones using the second random-phase approximation (SRPA)~\cite{Toh12a}, where they observed a complex microscopic structure of the $1^-_1$ state. 

It has been realized that cluster states form an additional phase of nucleonic matter besides the mean-field-type states, see, {\it e.g.}, Ref.~\cite{Ebr14a} and references therein. Recent theoretical studies in the framework of the EDF method have once more shown that different cluster configurations can appear as excited configurations in nuclei ranging from $^{8}$Be up to $^{40}$Ca~\cite{Ebr14a}. The algebraic cluster model~\cite{Bij00a, Bij14a} has been very successful to describe these configurations in light nuclei. Here, two-body clusters are described in terms of the algebra of $U(4)$. In fact, this corresponds to the $sp$ version of the algebraic interacting boson model (IBM)~\cite{Iach87a, Kusn88a, Eng87a} and it has been proposed a long time ago that the $p$-boson is related to $\alpha$-cluster configurations as shown in Fig.~\ref{fig:e1shapes}\,{\bf a)}~\cite{Dal83a, Iach82a} and is expected to occur at the surface of heavier nuclei where the density is low, compare also Ref.~\cite{Roe14a}. 
 
In the present study, the {\it spdf} IBM~\cite{Iach87a, Kusn88a, Eng87a} has been adopted to systematically study the low-lying $J^{\pi}$=~1$^-$ states in the Nd isotopes and other rare-earth nuclei. This model has already been successfully applied to describe $E1$ excitations related to octupole degrees of freedom in the rare earths~\cite{Kusn88a, Babi05a} and actinides~\cite{Zamf01a, Zamf03a}. The proximity of the additional and still not interpreted 1$^-$ states up to 4\,MeV to the octupole 1$^-$ states indicates the need for full $spdf$ IBM calculations.

\begin{figure}[h,b]
\centering
\includegraphics[width=.98\linewidth]{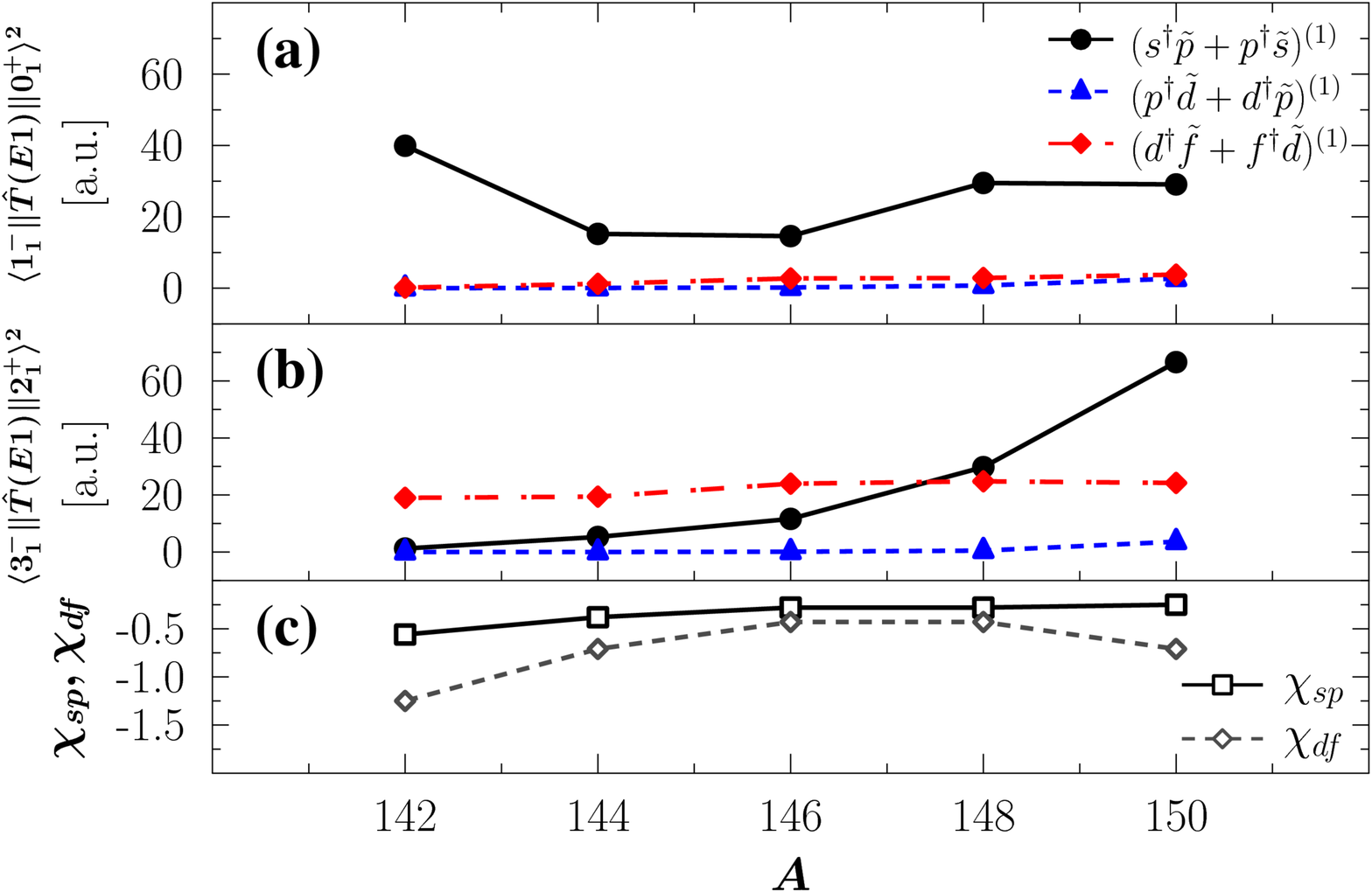}
\caption{\label{fig:Nd_ME} (Color online) $E1$ matrix element {\bf (a)} $\langle 1^-_1 \| \hat{T}(E1) \| 0^+_1 \rangle^2$ and {\bf (b)} $\langle 3^-_1 \| \hat{T}(E1) \| 2^+_1 \rangle^2$ for $^{142-150}$Nd. Shown are the bare contributions to the $spdf$ IBM $E1$ operator of Eq.~\ref{Eq:E1}. No fit parameters have been adjusted. {\bf (c)} Evolution of the fit parameters $\chi_{sp}$ and $\chi_{df}$ in Eq.\,\ref{Eq:E1} used to describe the experimental $E1$ strength.}
\end{figure}

The following Hamiltonian has been chosen, which is a natural extension of the $\hat{H}_{\mathrm{sd}}$ Hamiltonian~\cite{Kusn88a}:

\begin{align}
\label{eq:hamiltonian}
\hat{H}_{spdf}~=~&\epsilon_{d}\hat{n}_{d} + \epsilon_{p}\hat{n}_{p} + \epsilon_{f}\hat{n}_{f} - \kappa \hat{Q}_{spdf} \cdot \hat{Q}_{spdf} \nonumber
\\
&+ a_3 \left\lbrack \left(\hat{d}^{\dagger}\tilde{d}\right)^{(3)} \cdot \left(\hat{d}^{\dagger}\tilde{d}\right)^{(3)} \right\rbrack^{(0)},
\end{align}

with $\epsilon_d$, $\epsilon_p$, and $\epsilon_f$ being the boson energies and $\hat{n}_d$, $\hat{n}_p$, and $\hat{n}_f$ the number operators, respectively. The quadrupole interaction strength $\kappa$ of the quadrupole operator $\hat{Q}_{spdf}$ is the same for describing positive- and negative-parity states simultaneously. To account for experimental anharmonicities, the O(5) Casimir operator~\cite{Pasc10a, Pasc12a} has also been added. $a_3$ corresponds to the strength of this $l$=~3 interaction. We note explicitly that the choice of $\hat{H}_{spdf}$ describes $sd$ and $pf$ states separately and does not mix positive- and negative-parity boson states; i.e. besides the negative-parity boson energies all parameters are determined according to the signatures of well-established positive-parity collective states. For positive-parity states, these included the energies of the $2^+_1$, $4^+_1$, $0^+_2$, and $2^+_{\gamma}$ states as well as their reduced transition strengths and $\gamma$-decay branching ratios. For this approach see, {\it e.g.}, Ref.~\cite{McCut04a}. The $p$- and $f$-boson energies were fixed to describe the energies of the $1^-_1$ and $3^-_1$ states. To describe $E1$ transitions, the full one-body $E1$ operator was adopted: 

\begin{align}
\label{Eq:E1}
\hat{T}(E1) = &e_1 \lbrack \chi_{sp} (s^{\dagger} \tilde{p} + p^{\dagger} \tilde{s})^{(1)} + (p^{\dagger} \tilde{d} + d^{\dagger} \tilde{p})^{(1)} \nonumber \\
&+ \chi_{df} (d^{\dagger} \tilde{f} + f^{\dagger} \tilde{d})^{(1)} \rbrack,
\end{align}

and its parameters were smoothly varied, see Fig.~\ref{fig:Nd_ME}\,{\bf (c)}, to reproduce the reduced transition strengths related to the decays of the $1^-_1$ and $3^-_1$ states to the yrast positive-parity states.

\begin{figure*}[t]
\centering
\includegraphics[width=.9\linewidth]{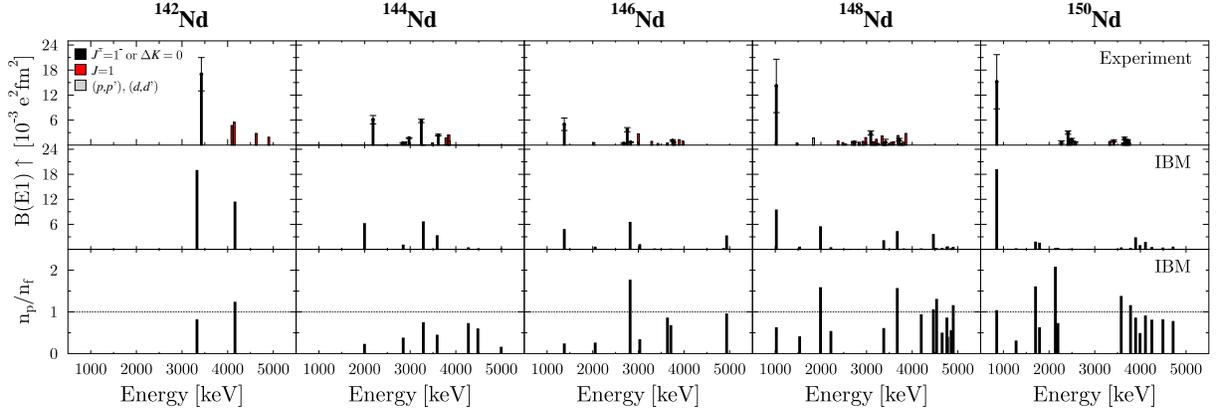}
\caption{\label{fig:Nd_E1} (Color online) {\it E1} distribution in ${}^{144-150}$Nd. Top panel: Experimentally firmly assigned $J^{\pi}$=~1$^-$ states or assigned $\Delta K$=~0 ground-state transitions are marked in black~\cite{ENSDF, Pitz90a, Fried92a, Eck97a}. States with $J$=1 assignment but no parity and no $\Delta K$=0 assignment are marked in red~\cite{Pitz90a}. Dipole states measured in Ref.~\cite{Pigna93a} by means of $(p,p')$ and $(d,d')$ reactions are presented in gray. Midpanel: $B(E1) \uparrow$ strengths and bottom panel: $n_p/n_f$ ratios predicted by the {\it spdf} IBM calculations, respectively.}
\end{figure*}

The calculations for the Nd isotopes are compared to the experimental results in Fig.~\ref{fig:Nd_E1} and Table~\ref{tab:e1}. Both the evolution of the excitation energy of the $1^-_1$ state as well as the $B(E1)\uparrow$ strength are nicely described. Typically, this state has been interpreted as the $K$=~0 projection of the one-octupole phonon excitation in deformed nuclei and as a candidate for the $J^{\pi}$=~1$^-$ member of the $(2^+_1 \otimes 3^-_1)$ quintuplet in vibrational and spherical nuclei~\cite{Knei06a}. Indeed, when identifying the basis states by means of $|\lbrack n_s \rbrack \lbrack n_p \rbrack \lbrack n_d \rbrack \lbrack n_f \rbrack \rangle$~\cite{Pasc12a}, then the $1^-_1$ in ${}^{144}$Nd is found to be dominated with 55~$\%$ by the $|4 0 1 1 \rangle$ configuration. This corresponds to the two-phonon $1^-$ state~\cite{Rob94a}. In addition, several $J$=~1 states are observed up to $\sim$~4~MeV in experiment, for which negative parity has been assigned either by means of their $K$ quantum number assignment based on their $\gamma$-decay properties~\cite{Pitz90a}, i.e. $\Delta K$=~0, or by direct parity measurements~\cite{Fried92a, Fried94a, Eck97a}. These states are marked black in the top panel of Fig.~\ref{fig:Nd_E1}. For the other $J$=~1 states, the experimental situation without parity assignment is more complex since also the scissors mode contributes to the dipole distribution in deformed nuclei at $E_x \thickapprox$~3~MeV~\cite{Heyd10a}. Based on the present experimental data, it cannot be excluded that some of the $J$=~1 states do have positive parity. These states are marked in red. Furthermore, $J^{\pi}$=~1$^-$ candidates observed in inelastic scattering experiments are marked in gray~\cite{Pigna93a}. Their scattering cross section has been scaled with respect to the observed one for the $1^-_1$ state and its respective $B(E1)$ value. Most likely, they have not been observed in the $(\gamma,\gamma')$ experiments due to the high background at low energies usually present in NRF experiments with bremsstrahlung. Keeping these limitations in mind, a good agreement between experiment and the present calculations is recognized for the additional $E1$ strength.

\begin{table}[h]
\caption{Excitation energy and $E1$ strength for the $1^-_1$ state as well as centroid energy and summed $E1$ strength for the remaining $1^-$ states in the Nd isotopes.}
\label{tab:e1}
\begin{ruledtabular}
\begin{tabular}{ccccc}
$A$ & $E_{1^-_1}$ & $B(E1)_{1^-_1} \uparrow$ & $E_{1^-_{w/o~1^-_1}}$ & $\sum B(E1)_{w/o~1^-_1} \uparrow$ \\
 & [MeV] & [$10^{-3}\,\mathrm{e^2fm^2}$] & [MeV] & [$10^{-3}\,\mathrm{e^2fm^2}$] \\
\hline
\multicolumn{5}{c}{Experiment} \\
142 & 3.4 & 17(4) & - & -\\
144 & 2.19 & 6.1(10) & 3.5 & 19.5(6)\footnote{10.4(5) for firm $J^{\pi}=1^-$ states and $\Delta K = 0$ assignment.} \\
146 & 1.38 & 5(2) & 3.3 & 14(2) \\
148 & 1.02 & 14(6) & 3.3 & 30(2)\footnote{17.6(13) for $J=1$ states with $K$ assignment.} \\
150 & 0.85 & 15(7) & 3.2 & 18.3(11)\footnote{11.4(8) for firm $J^{\pi}=1^-$ states and $\Delta K = 0$ assignment.} \\
\multicolumn{5}{c}{$spdf$ IBM} \\
142 & 3.3 & 19 & - & - \\
144 & 2.00 & 6.7 & 3.4 & 11.5 \\
146 & 1.38 & 5 & 3.4 & 12 \\
148 & 1.02 & 9 & 3.3 & 19 \\
150 & 0.85 & 19 & 3.3 & 11.2
\end{tabular}
\end{ruledtabular}
\end{table}

\begin{figure*}[t]
\centering
\includegraphics[width=.98\linewidth]{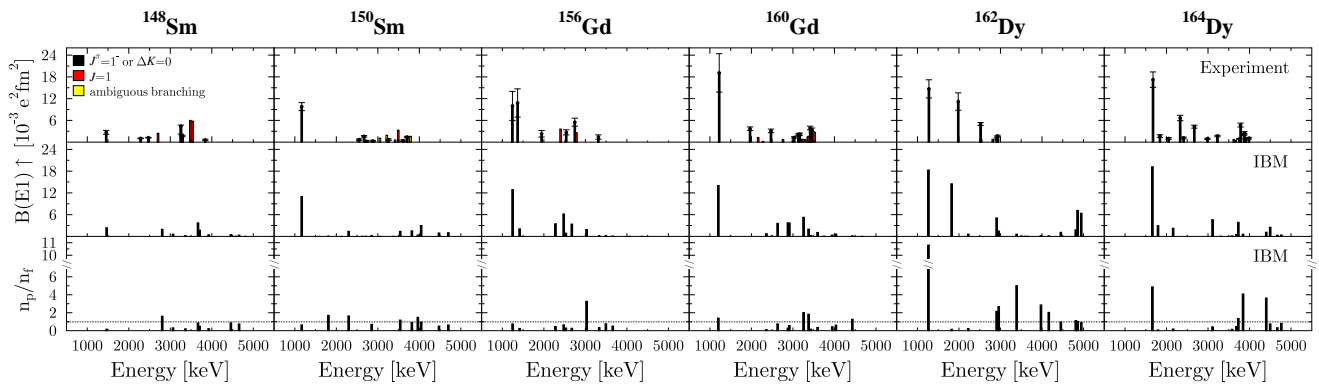}
\caption{\label{fig:Others_E1} (Color online) Same as Fig.~\ref{fig:Nd_E1} but for ${}^{148,150}$Sm, ${}^{156,160}$Gd, and ${}^{162,164}$Dy. In addition, states with an experimentally ambiguous $\gamma$-decay branching are marked in yellow.}
\end{figure*} 

To further study the structure of the states, it is necessary to look beyond the strength distribution. The lowest panel of Fig.~\ref{fig:Nd_E1} shows the $p$- and $f$-boson fraction $n_p/n_f$. For illustrative purposes, only states are shown having a $B(E1)$ larger than 0.1~$\times~10^{-3}$~e$^2$fm$^2$. In all Nd isotopes, states are observed which show an enhanced $p$-boson content, i.e. $n_p/n_f > 1$, and which make up for a great part of the enhanced $E1$ strength.

Another remarkable observation is the stability of the centroid energy and summed strength of this additional $E1$ strength which is located around 3.4~MeV, see Table~\ref{tab:e1}. This automatically raises the question if and how this strength is connected to the proposed $(2^+ \otimes 3^-)_{1^-}$ state in ${}^{142}$Nd at 3.4~MeV with $B(E1) \uparrow \, = 17(4) \cdot 10^{-3} \, \mathrm{e^2fm^2}$~\cite{Wilh98a}. This structure assignment was based upon two facts: (1) $E_{1^-} \thickapprox E_{2^+}+E_{3^-}$ and (2) $B(E2;1^- \rightarrow 3^-_1) \thickapprox B(E2;2^+_1 \rightarrow 0^+_1)$~\cite{Wilh98a}, supposed key signatures of a QOC state. However, a final assignment also requires another proof, namely, that $B(E3;1^- \rightarrow 2^+_1) \thickapprox B(E3;3^-_1 \rightarrow 0^+_1)$. The inspection of Fig.~\ref{fig:Nd_E1} shows that the $1^-_1$ state in $^{142}$Nd has a non-neglibile $p$-boson content ($n_p/n_f$=0.8). The  $|5 1 0 0 \rangle$ configuration accounts for 28\,$\%$ of the total wave function ($<$10\,$\%$ in $^{144}$Nd), while  $|4 0 1 1 \rangle$ contributes with 36\,$\%$. 

Furthermore, the  $(s^{\dagger} \tilde{p} + p^{\dagger} \tilde{s})^{(1)}$ contribution to the $\langle 1^-_1 \| \hat{T}(E1) \| 0^+_1 \rangle^2$ matrix element in Fig.~\ref{fig:Nd_ME}\,{\bf (a)} clearly resembles the experimentally determined parabolic behavior  of the $E1$ strength, see, {\it e.g.}, Ref.~\cite{Frans98a} and references therein. Realizing the $sp$-boson space as the bosonic manifestation of $U(4)$, which was proposed to describe two-body clusters~\cite{Bij00a, Bij14a}, no complex mixing with GDR components is needed.  As expected from macroscopic considerations, the other matrix elements nearly vanish at shell closure and cannot explain the $E1$ strength increase at shell closure. The $p$-boson structure also explains the near equality of the two mentioned $B(E2)$ values since they correspond to $E2$ transitions of the type $s^{\dagger}\tilde{d}$ and $f^{\dagger}\tilde{p}$, which lead to comparable strength when considering the $spdf$-IBM $E2$ operator~\cite{Iach87a}. We also note that comparable $E1$ matrix elements, found in this Letter, can explain the suggested empirical correlation between the $B(E1;1^-_1 \rightarrow 0^+_1)$ and $B(E1; 3^-_1 \rightarrow 2^+_1)$ values in vibrational nuclei~\cite{Piet99a}, see Fig.\,\ref{fig:Nd_ME}\,{\bf (a)} and {\bf (b)}.

To show the general importance of the $p$ boson and the appearance of $p$-boson states all over the rare-earth region, we have also studied ${}^{148,150}$Sm, ${}^{156,160}$Gd, and ${}^{162,164}$Dy. The parameters for positive-parity states of the Dy isotopes were taken from Ref.~\cite{McCut04a}. The results are shown in Fig.~\ref{fig:Others_E1}. As in the Nd isotopes, the general agreement with the experimentally observed $E1$ distribution is good. The similarities between $^{148,150}$Sm and $^{146,148}$Nd are in fact striking. Very strong $p$-boson states are observed in the Dy isotopes.

The $\alpha$-clustering mode, i.e. the $p$-boson mode, corresponds to 4-QP configurations, which were also proposed in the framework of the EDF approach to obtain a better description of the 1$^-$ states~\cite{Toh12a}. Calculations of this kind have not been performed yet but the importance of the particle-particle interaction for semimagic nuclei has been pointed out in Ref.~\cite{Grin94a}. In Ref.~\cite{Jol04a} it was shown that 2-QP components had to be admixed to the $1^-_1$ states to describe the $E1$ strength, even though deviations were found for the Nd and Sm isotopes. It was speculated that these admixtures could be mimicked by the $p$-boson. The clear evolution of $\epsilon_p$ from 1.5\,MeV in $^{150}$Nd to 4\,MeV in $^{142}$Nd and, especially, the independent evolution of the $\langle 1^-_1 \| \hat{T}(E1) \| 0^+_1 \rangle^2$ and $\langle 3^-_1 \| \hat{T}(E1) \| 2^+_1 \rangle^2$ matrix elements as well as the non-negligible $p$-boson contribution to the norm of the wave function, in contrast to Ref.~\cite{Jol04a}, are more likely to point to an evolution of a configuration complementary to the mean-field-like states and might provide a new understanding of the origin of low-lying $E1$ strength in nuclei.

In conclusion, the experimental $E1$ strength distribution below 4\,MeV suggests the presence of a new collective dipole mode different from the octupole mode in rare-earth nuclei~\cite{Frans98a}, i.e. the $\alpha$-clustering mode occuring at the surface of nuclei just above magic numbers~\cite{Iach82a, Iach85a}. In this Letter, we have presented, for the first time, a systematic study of the $E1$ strength observed in rare-earth nuclei using the $spdf$ IBM. In agreement with experiment, several enhanced $E1$ transitions are observed. The new class of excitations is closely related to the $p$ boson. Furthermore, an alternative interpretation of the peculiar $B(E1)$ increase at shell closure might be possible in terms of the fundamental $p$-boson mode, i.e., $\alpha$ clustering~\cite{Iach82a, Dal83a}. Our studies suggest a close connection between $E1$ strength and $\alpha$ clustering~\cite{Iach85a}, i.e., an additional isospin-symmetry breaking component, in heavier nuclei and might hint at the general occurrence of this mode in nuclei. Microscopic calculations looking into the details of the wave functions as well as experiments sensitive to $\alpha$ structures, $e.g.$, $(d,{}^{6}Li)$, $({}^{6}Li,d)$, $(\vec{\gamma},\gamma)$, and $(e,e')$ reactions, are asked for. Furthermore, the $(\gamma,\gamma')$ measurements should be extended up to the particle-emission threshold to study the complete low-lying $E1$ strength with evolving deformation.

\begin{acknowledgments}
The experimental $(\gamma,\gamma')$ data have been obtained in a very close collaboration of the Universities of Stuttgart, Darmstadt, and Köln. We thank all colleagues from this collaboration. Discussions with V. Derya, A. Hennig, and P. von Brentano are highly appreciated. This work was supported by the DFG (ZI 510/4-2) and in part by the U.S. Department of Energy Grant No. De-FG-02-91ER-40608. M.S. is supported by the Bonn-Cologne Graduate School of Physics and Astronomy.
\end{acknowledgments}

\bibliography{LIR}

\end{document}